\documentclass[english,aps,twocolumn]{revtex4}
\usepackage[T1]{fontenc}
\usepackage[latin9]{inputenc}
\usepackage{caption}
\usepackage{amsmath}
\usepackage{amssymb}
\usepackage{graphicx}
\usepackage{epstopdf}

\makeatletter
\@ifundefined{textcolor}{}
{%
 \definecolor{BLACK}{gray}{0}
 \definecolor{WHITE}{gray}{1}
 \definecolor{RED}{rgb}{1,0,0}
 \definecolor{GREEN}{rgb}{0,1,0}
 \definecolor{BLUE}{rgb}{0,0,1}
 \definecolor{CYAN}{cmyk}{1,0,0,0}
 \definecolor{MAGENTA}{cmyk}{0,1,0,0}
 \definecolor{YELLOW}{cmyk}{0,0,1,0}
}

\makeatother
\usepackage{babel}
\begin{document}

\title{Colossal proximity effect in  a superconducting triplet spin valve based on halfmetallic ferromagnetic CrO$_2$}


\author{A. Singh, S. Voltan, K. Lahabi \& J. Aarts}
\affiliation{Huygens - Kamerlingh Onnes Laboratory, Leiden University, P.O. Box 9504, 2300 RA Leiden, The Netherlands.}

\date{today}

\maketitle
\selectlanguage{english}%

\maketitle
\selectlanguage{english}%

\textbf{Ferromagnets can sustain supercurrents through the formation of equal spin triplet Cooper pairs and the
mechanism of odd-frequency pairing. Since such pairs are not broken by the exchange energy of the ferromagnet,
superconducting triplet correlations are long-ranged and spin-polarized, with promises for superconducting spintronics
devices. The main challenge is to understand how triplets are generated at the superconductor (S)/ ferromagnet (F)
interface. Here we use the concept of a so-called triplet spin valve (TSV) to investigate the conversion of singlets in
a conventional superconductor to triplets in the halfmetallic ferromagnet CrO$_2$. TSV's are composed of two
ferromagnetic layers (separated by a thin normal metal (N) layer) and a superconductor (F$_1$/N/F$_2$/S). The package
F$_1$/N/F$_2$ generates triplets in F$_1$ when the magnetization directions of the F$_{1,2}$-layers are not collinear.
This drains singlet pairs from the S-layer, and triplet generation is therefore signalled by a decrease of the critical
temperature $T_c$. Recently, experiments with TSV's were reported with Co draining layers, using in-plane fields, and
finding $T_c$-shifts up to 100~mK. Using CrO$_2$ instead of Co and rotating a magnetic field from in-plane to
out-of-plane, we find strong $T_c$ variations of almost a Kelvin up to fields of the order of a Tesla. Such strong
drainage is consistent with the large lengths over which supercurrents can flow in CrO$_2$, which are significantly
larger than in conventional ferromagnets. Our results point to the special interest of halfmetals for superconducting
spintronics. }

Combining superconductors (S) and ferromagnets (F) offers the opportunity to create a new class of superconducting
spintronic devices \cite{fominov10,feofanov10,linder13}. In such S/F hybrids a long range spin polarized triplet
supercurrent can be generated by converting Cooper pairs from the singlet to triplet state via spin mixing and spin
rotation at the S-F interface, which requires the presence of magnetic inhomogeneity
\cite{bergeret01,kadigr01,bergeret03,houzet07,eschrig08}. Recently, it was shown that long range supercurrents could be
engineered in S/F/S Josephson junctions by inserting an extra ferromagnetic layer between the superconductor and the
central F-layer \cite{khaire10,robinson10,khasawneh12,anwar12}. Still, quantitative understanding of the conversion
process is mostly lacking since the spin activity of the interface is not a measurable parameter. Absolute values of
the supercurrent are not easily predictable, which was illustrated clearly in recent work of Klose {\it et al}
\cite{klose12}, where supercurrents in a Co-based Josephson junction could be increased more than an order of magnitude
by manipulating the magnetization directions of F$_1$ and F$_2$. For acquiring such understanding, a Josephson junction
has the disadvantage that it contains two sets of interfaces, which may not have the same amount of spin activity or
even transparency. In this sense a triplet spin valve (TSV), pictorially sketched in Fig.~1a, is a simpler device. It
can be thought of as half of the Josephson junction, utilizing the same layer package F$_1$/N/F$_2$/S. The S-layer is
chosen not too thick, so that drainage of Cooper pairs through triplet conversion is reflected in the change of $T_c$
of the stack, F$_2$ is thin in order to take part in the spin mixing, but not to break Cooper pairs, and F$_1$ is the
drainage layer, which can be infinitely thick. By changing the relative magnetization directions of F$_1$ and F$_2$ the
triplet pair generation is varied and thereby the amount of singlet pairs which is converted, making the operation a
field-controlled proximity effect. When F$_1$ and F$_2$ are orthogonal, triplet generation is maximum and $T_c$ should
be minimum. The performance of an TSV can be gauged by the extent to which $T_c$ decreases, and, as will be shown
below, also interface transparency can be explicitly addressed.
\\
There are several recent experimental results on TSVs, which use standard ferromagnets (Fe, Co, Ni) and their alloys as
spin mixers and drainage layers \cite{leksin12,wang14,banerjee14,jara14,flokstra14}. In all cases magnetic anisotropy
or an antiferromagnetic pinning layer was used to reliably control the relative magnetization directions, always in the
plane of the films. The maximum suppression of T$_{c}$ achieved in such devices ranged from $120$ mK (for a thin
ballistic Co drainage layer)\cite{wang14} to $20$ mK (for diffusive TSVs)\cite{leksin12,jara14,flokstra14}. Our
experiments are different in two important aspects. One is the use of the halfmetallic ferromagnet CrO$_{2}$ as the
drainage layer. The other is that we vary the field from in-plane to out-of-plane. A disadvantage is that the field
rotation also changes the critical field of the superconductor, which is not the case when fields are rotated in plane.
\begin{figure*}[]
\captionsetup{width=\textwidth}
\begin{tabular}{lc}
\includegraphics[width=7cm]{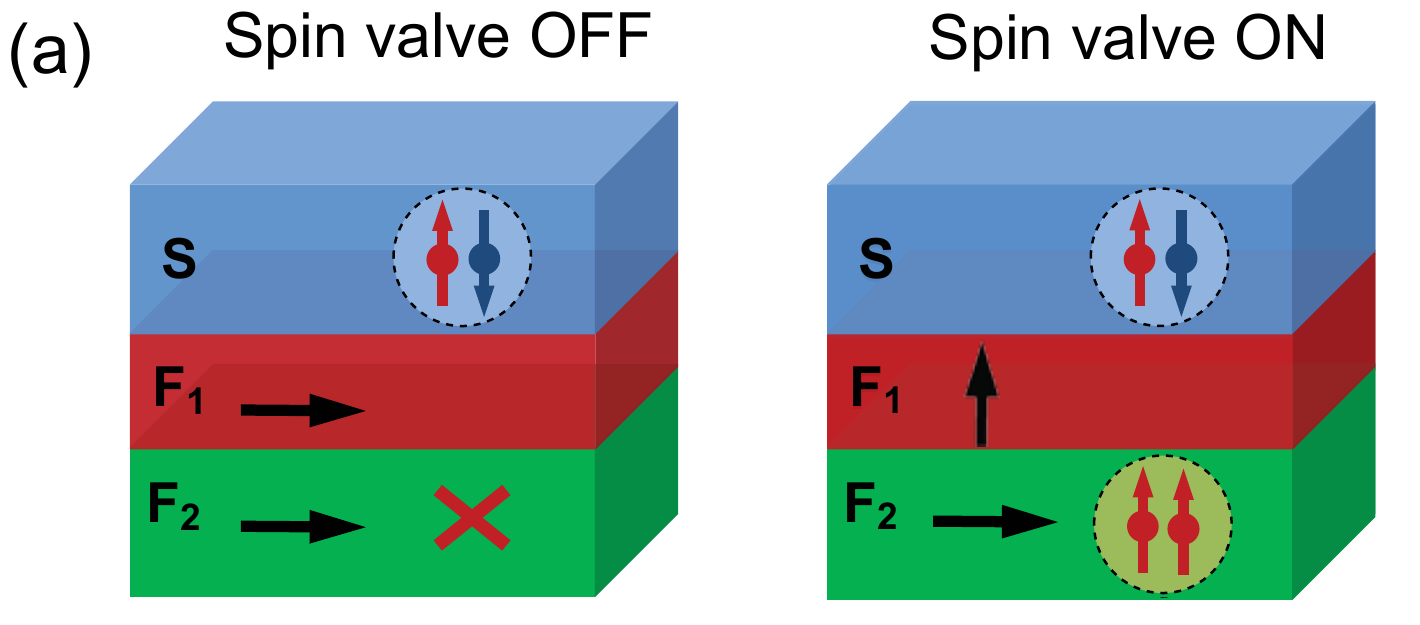}  \hspace{1cm} &
\includegraphics[width=9cm]{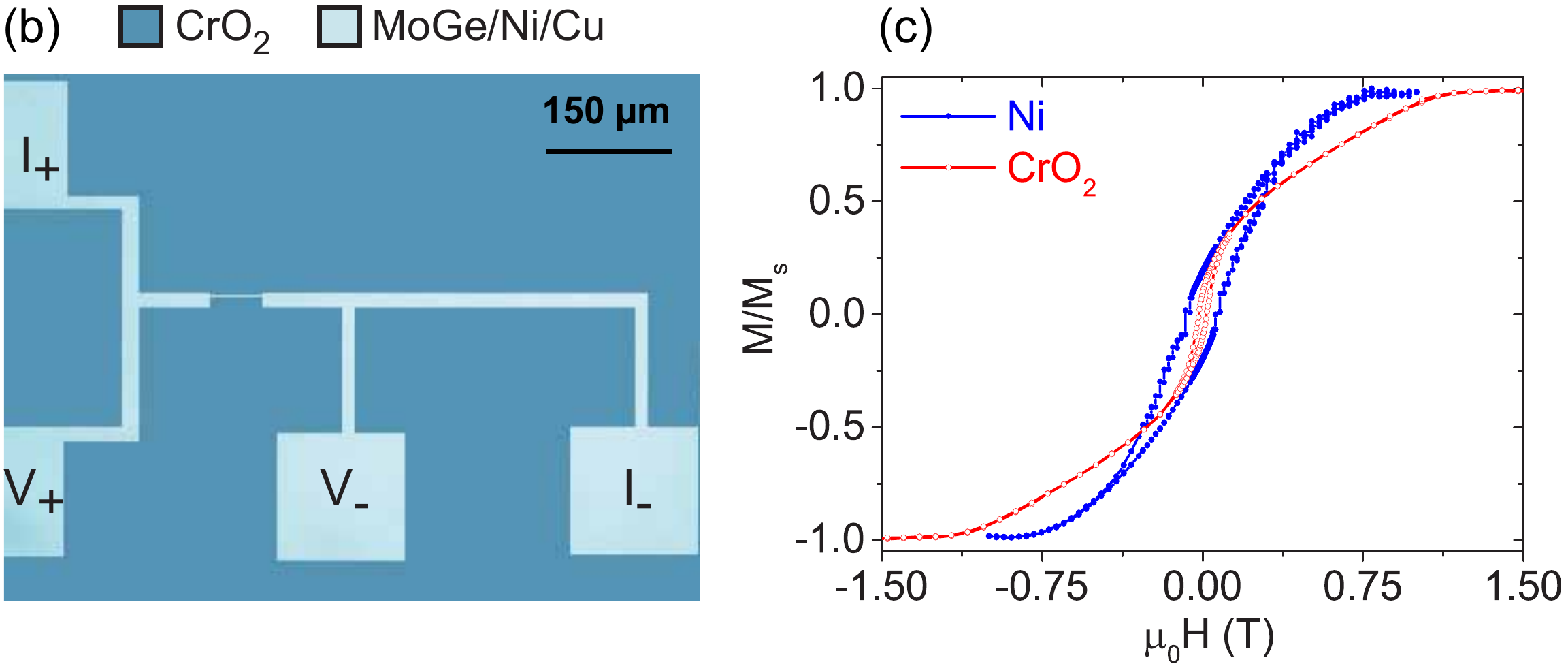}
\end{tabular}
\caption[width=1\textwidth]{$\textbf{Device structure and magnetic characterization of the ferromagnetic layers in the
spin triplet valve.}$ $\bf{a}$, Working principle of a Triplet Spin Valve (TSV)with a half metallic ferromagnet; the
TSV is off (on) when the magnetizations of F$_1$ and F$_2$ are collinear (non-collinear), with a maximum effect when
they are orthogonal. $\bf{b}$, Optical micrograph of a typical TSV where a MoGe(d$_{s}$)/Ni($1.5$ nm)/Cu ($5$ nm)
trilayer bridge of $10 \mu$m width was patterned on a $100$ nm thin film of CrO${_2}$. $\bf{c}$, Magnetization
hysteresis loops for CrO${_2}$(100 nm) and a multilayer (Ni(1.5)/Cu(10))$_{11}$/Ni(3)/Cu(10) measured with the magnetic
field perpendicular to the sample plane. The magnetization $M$ is normalised on the saturation magnetization $M_s$,
which was 6.8~$\times 10^5$ A/m for the CrO$_2$ film and 2.2~$\times 10^5$ A/m for the Cu/Ni multilayer.}
\end{figure*}
We solve this by comparing our TSV with stacks where the $F_2$ layer is absent, as well as with the simple S-layer. In
this way we find that spin valve effects are present up to fields of Tesla's, and they are remarkably large, with a
suppression in T$_{c}$ as high as $800$~mK at 2~T. The origin of this significant variation probably lies in the fact
that CrO$_{2}$ is $100$$\%$ spin polarized and strongly supports triplet correlations. \\
Our TSV is made of amorphous MoGe, Ni, Cu and CrO$_{2}$ as the S, F$_1$, N and F$_2$ layers, respectively. The Cu layer
is required to magnetically decouple the mixer and drainage layers. The experimental device consists of a CrO$_{2}$
film grown a TiO$_2$ substrate by chemical vapor deposition, on top of which a $10 \mu$m wide MoGe/Ni/Cu trilayer
bridge was deposited using sputtering and lift-off \cite{anwar10}. Prior to the trilayer deposition, the top surface of
CrO$_{2}$ was cleaned with an Argon ion plasma to remove the thin insulating Cr$_{2}$O$_{3}$ barrier which is prone to
form at the end of the deposition process. An optical micrograph of one such device is shown in Fig.~1b. To
characterize the magnetic properties of F$_{1,2}$ layers, their hysteresis loops (magnetization $M$ versus applied
field $H_a$) were measured using SQUID magnetometry in the out-of-plane configuration. Instead of a single Ni layer of
1.5~nm, we used a multilayer (Ni(1.5)/Cu(10))$_{11}$/Ni(3)/Cu(10) in order to boost the signal. The data are given in
Fig.~1c and show that in both layers the rotation of the magnetization requires a field of order of a Tesla.
Electrical measurements were performed in a four probe configuration in a physical properties measurement systems
(PPMS). For angle resolved magnetotransport measurements the magnetic field ($H$) was rotated in a plane normal to the
sample.  In this geometry, when $\theta = 0^{\circ}$  the field is aligned with the current density ($j$), and $\theta
= 90^{\circ}$ corresponds to the out-of-plane applied field as outlined in Fig.~2a.
Fig.~2b (upper panels) shows $R$(T)-curves at different angles of the magnetic field for two TSVs consisting of
MoGe(d$_S$/Ni($1.5$ nm)/Cu($5$ nm)/CrO$_{2}$ ($100$ nm), with two different values of the MoGe thickness d$_S$ = 25~nm
(called TSV25) and d$_S$ = 50~nm (called TSV50), and for fixed magnetic fields of 0.25~T and 0.5~T. We extract an
operational parameter T$_{50\%}$ (for a brief discussion of this choice, see the supplementary information) which is
the temperature where the resistance has decreased to $50\%$ of the normal resistance value. The variation of
T$_{50\%}$ with $\theta$ is called $\delta$T$_{50\%} \equiv$ T$_{50\%}(0^{o})$ - T$_{50\%}(\theta)$. The lower panels
show $\delta$T$_{50\%}$ as function of $\theta$, and the curves clearly exhibit a maximum when the field is normal to
the plane. Further points to note are (i) the large values of the change, of about 550~mK and 650~mK for TSV50 in
0.25~T and 0.5~T, respectively, and 750~mK for TSV25 in 0.5~T;
(ii) the significantly larger value of the normal state resistance for TSV25; (iii) the sharp peak in resistance which
in parallel field occurs at the onset of superconductivity and which smears out and disappears when rotating the field.
\\
\begin{figure*}[]
\captionsetup{width=1\textwidth}
\begin{tabular}{lc}
\includegraphics[width=4cm]{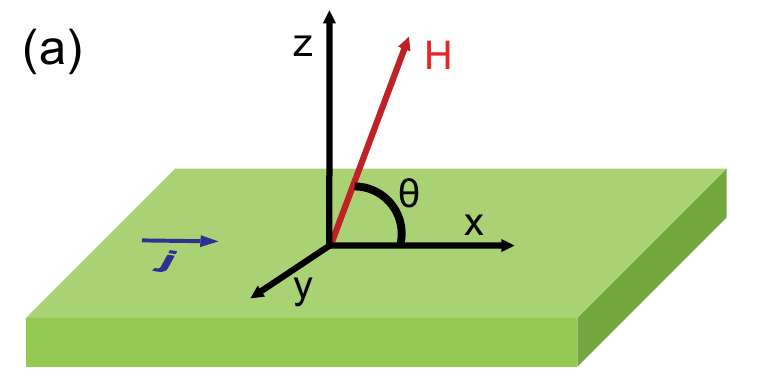}  \hspace{1cm} &
\includegraphics[width=12cm]{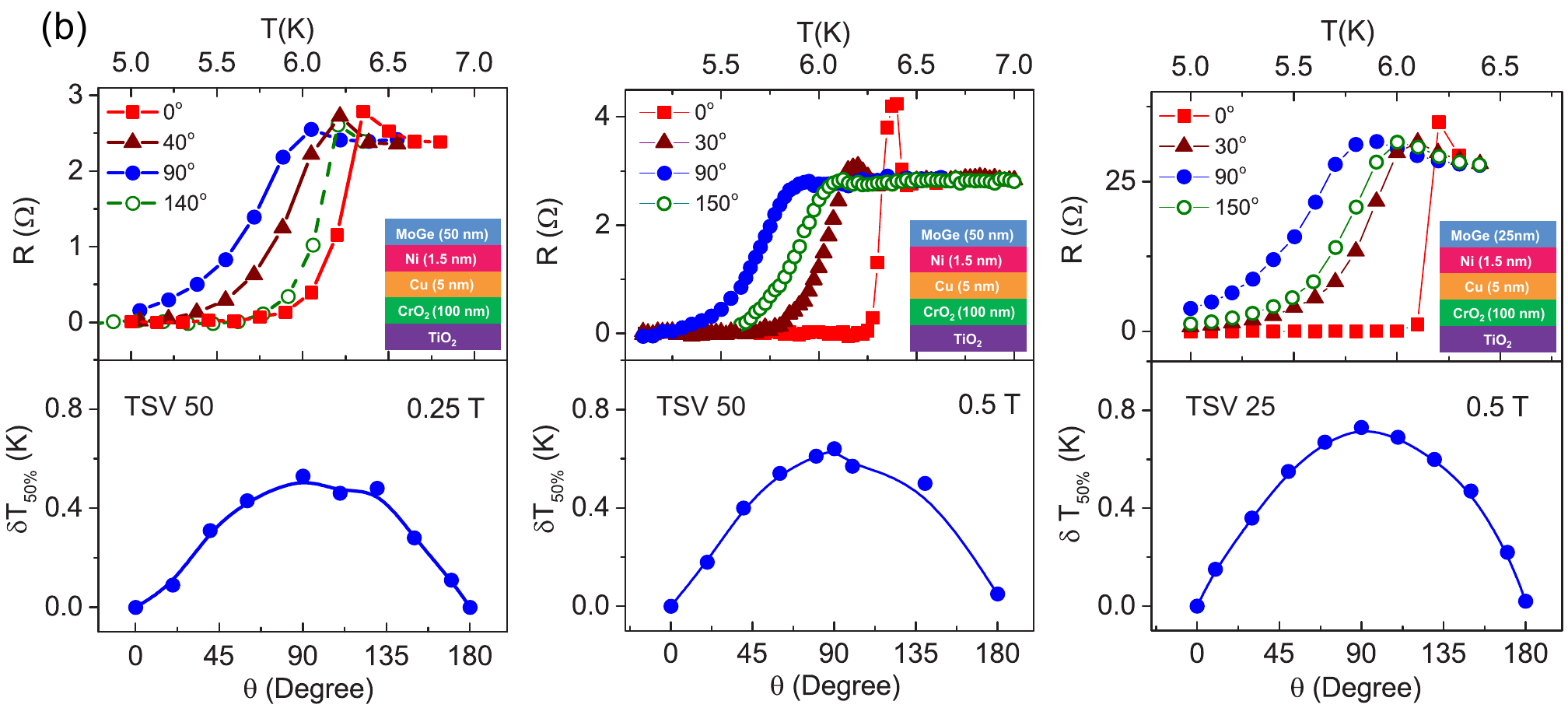}
\end{tabular}
\caption[width=1\textwidth]{\textbf{Dependence of the critical temperature of the TSVs on the direction of the applied
field.} $\bf{a}$, Coordinate system used in angle dependent magnetotransport measurements, showing the direction of the
current $j$, the applied field $H_a$ and the angle $\theta$ between them; $\bf{b}$, Spin valve effect in the two spin
valves MoGe(d$_{s}$)/Ni($1.5$ nm)/Cu($5$ nm)/CrO$_{2}$($100$ nm) with d$_{s}$ = 50~nm (TSV50; left and middle), and
d$_{s}$ = 25~nm (TSV25; right). Upper panels: resistive transitions for different $\theta$ as indicated. Lower panels:
variation of $\delta T_{50\%}$ = T$_{50\%},(0^{o})$ - T$_{50\%},(\theta)$ as a function of $\theta$ at 0.25~T (TSV50)
and 0.5~T (TSV50, TSV25), where $T_{50\%}$ is the temperature where the normal state resistance has decreased by 50~\%.
Note that a peak appears in the transition curves for measurements at $\theta~=~0^{o}$.}
\end{figure*}
In order to discuss the first point we have to put the data in perspective. The superconductor itself will show a
T$_{50\%}(\theta)$ variation, because the transition in parallel field is due to the onset of surface
superconductivity, which is at a higher field and temperature than the transition in perpendicular critical field. The
change from surface to bulk effects also raises concerns about going from a vortex-free configuration to one where
vortex flow may play a role. These issues are resolved by a straightforward comparison with the behavior of a single
MoGe layer, for which we take a thickness of 50~nm. Stray fields of mixer and drainage layer may also play a role, and
therefore we compare with devices of MoGe(50)/Ni(1.5)/Cu(5) and MoGe(50)/Cu(5)/CrO$_{2}$ as well. Fig.~3a shows the
transition curves of these devices at 0.25~T for in-plane and out-of-plane configurations. All have comparable
$\delta$T$_{50\%}$. In Fig.~3b values of $\delta T_{50\%}$ for the different data sets are compared, again at a field
of 0.25~T. It can be clearly seen that the variation in the TSV is significantly larger than in the other devices.
Fig.~3c shows the variation of $\delta$T$_{50\%,max}$ = T$_{50\%}(0^{\circ})$-T$_{50\%}(90^{\circ}$) as a function of
the applied field for an isolated MoGe film  and a TSV. In both cases $\delta T_{50\%,max}$ increases monotonically
with the magnetic field up to 2~T. The shaded area in Fig.~3c solely corresponds to the effect of triplet generation
which can suppress $T_{50\%}$ by as much as 800~mK. We find TSV effects over a wide range of magnetic fields. This was
not the case for previous TSVs measured in an in-plane configuration, where the maximum field of operation was limited
to $0.2-0.3$ T \cite{flokstra14}. Robust proximity effects were also observed in CrO$_{2}$ based Josephson junctions
\cite{keizer06,anwar10,anwar12} where critical currents in various configurations were observed up to the Tesla range.
Slightly puzzling is that $\delta $T$_{50\%,max}$ continues to increase well above the fields where saturation of both
mixer and drain layer have been achieved and both magnets are assumed to be collinear. We believe that this may be
caused by the presence of non-collinear magnetic moments pinned at the CrO$_{2}$ interface. Noting that the difference
between the TSV and the other stacks only lies in the insertion of an extra 1.5~nm layer of Ni, it seems reasonable to
conclude that, just as in the case of the Josephson junctions, the Ni layer is instrumental in generating triplets.
They are very efficiently drained by the CrO$_2$ layer which leads to the observed large spin valve
effects. \\
Turning to the larger value of the normal state resistance $R_N$ of TSV25, this can be used to probe the effects of the
bare interface transparency, which is a critical parameter in determining the strength of proximity effect, and much
studied in S/N and S/F hybrids \cite{fominov02,cirillo05}. In our devices the transparency of the interface between the
CrO$_2$ film and the Cu/Ni/MoGe stack is controlled by the Argon etching of the CrO$_{2}$ surface prior to the
deposition of the stack. The etching is a critical step in the fabrication, due to the fact that under-etching results
in only partial removal of an unwanted Cr$_{2}$O$_{3}$ layer while over-etching induces disorder at the surface of
CrO$_{2}$.
\begin{figure*}[]
\captionsetup{width=1\textwidth}
\begin{centering}
\includegraphics[width=12cm]{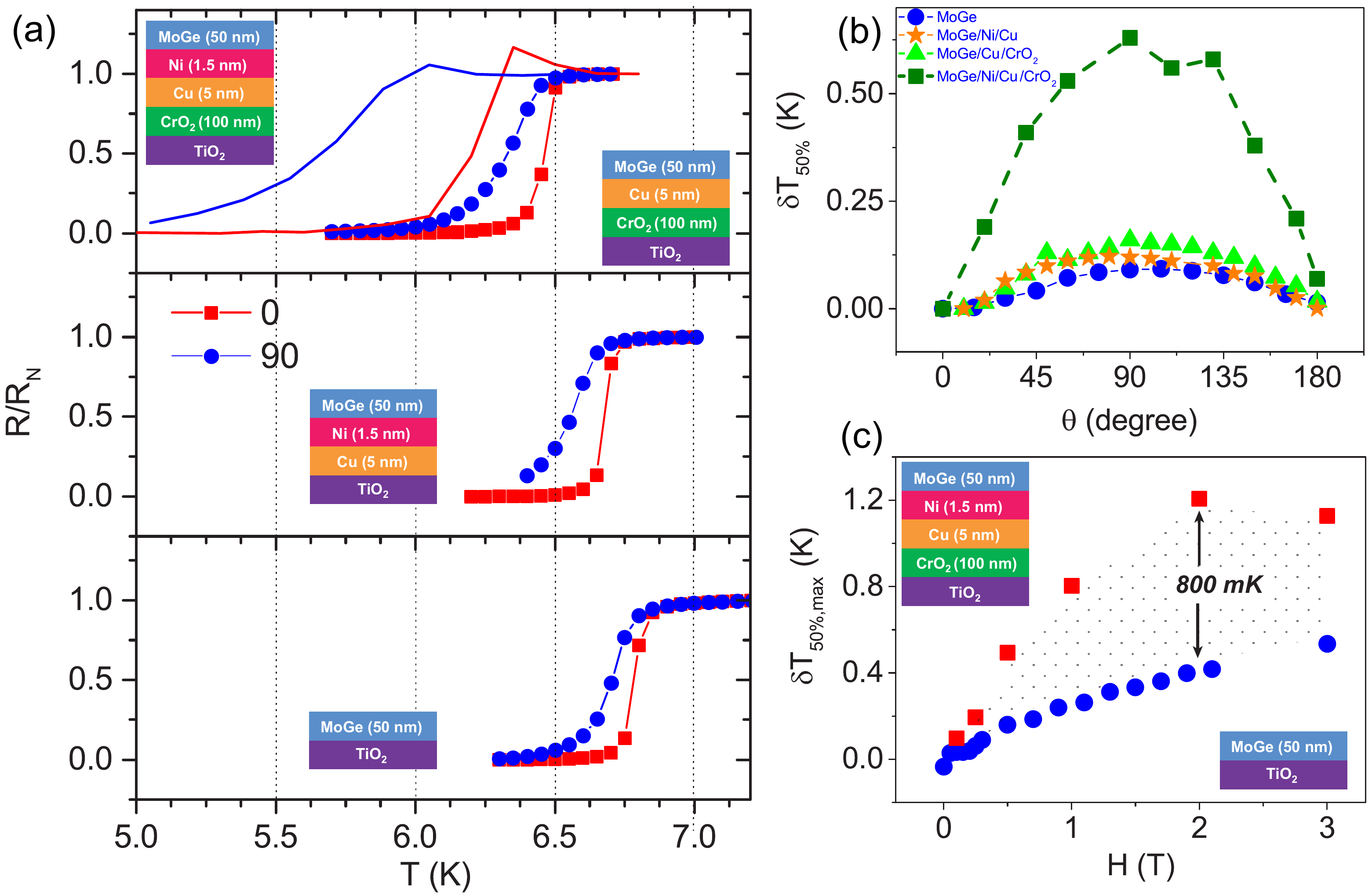}
\par\end{centering}
\caption{\textbf{Angular variation of the critical temperature for different non-triplet-generating layer combinations}
$\bf{a}$, Transition curves for MoGe(50)/Cu(5)/CrO$_{2}$ (top), MoGe(50)/Ni(1.5)/Cu(5) (middle), and MoGe(50) (bottom
panel) for $\theta = 0^{o}$ and $\theta = 90^{o}$ at $0.25$T. The top panel also shows the results of the spin valve
device TSV50 as drawn lines. $\bf{b}$ $\delta$ T$_{50\%}$ as function of $\theta$ for these layered devices and for
TSV50 at $0.25$T. $\bf{c}$, Variation of T$_{50\%,max}$ = T$_{50\%},(0^{o})$ - T$_{50\%},(90^{o})$ as function of
applied field for MoGe(50) and TSV50.}
\end{figure*}
We take advantage of this by making different devices on the same CrO$_2$ film using different etch times. For this the
film is covered with resist, a lift-off structure is written, the CrO$_2$ surface is etched for a certain amount of
time, and the stack is deposited. This process is repeated with different etch times. The transparency has a direct
influence on the normal resistance of the device, which in essence consists of a top N-layer (MoGe) of high resistance
and a bottom F-layer (CrO$_2$) of low resistance, with an interface resistance $R_B$ in between. With contacts on top,
$R_B$ is in series with the low-resistance bottom layer and has a measurable influence on $R_N$. Details are given in
the supplementary information. In Fig.~4 we plot $\delta$T$_{50\%,max}$ against 1/$R_N$ for both sets of devices
measured in 0.5~T. The performance of all  TSVs increases monotonically with decreasing $R_N$ and increasing barrier
transparency. Interface transparency also offers a natural explanation for the fact that TSV25 exhibit an effect
comparable to TSV50 with a thicker MoGe layer. According to basic proximity effect theory, the thinner layer should
show a stronger effect upon Cooper pair depletion, but as can be seen in Fig.~4, this is counteracted by the lower
interface transparency. In this respect it should also be remarked that TSV50 is surprisingly efficient when taking
into account that the S-layer thickness is about ten times the superconducting coherence length $\xi_S$, which for MoGe
is about 5~nm \cite{baarle03}. This again appears to be a consequence of the
$100\%$ spin polarized ferromagnet. \\
The final striking feature in our results is the observed characteristic peak in the transition curve of a TSV at a
finite in-plane field, which disappears when the field is rotated out of the plane (see Fig.~2b). As shown in Fig.~5a
for TSV50, the peak is not present in zero field, but then gradually appears at fields around 0.2~T, behavior which
appears consistently in all our devices. Here we have to speculate and attribute the effect to the normal reflection of
equal spin triplet Cooper pairs at the half metallic boundary, in a mechanism which has not yet been well described or
identified but requires magnetically misaligned moments at the CrO$_2$ interface. As pictorially shown in Fig.5b, we
assume that $m_s=0$ singlets are converted into $m_s=0$ triplets in the Ni/Cu sandwich, but that a triplet $m_s=1$
quantization axis is provided by the misaligned moments, which could be called an F' layer. When these $m_s=1$ triplets
encounter the CrO$_2$ bulk, they will partly be transmitted, but also partly reflected. The latter may result in the
breaking of the pair on the MoGe/Cu/Ni-side of the stack, resulting in quasiparticles with the same spin. This spin
accumulation raises the spin chemical potential ($\Delta\mu=\mu_{\uparrow}-\mu_{\downarrow}$) and results in additional
spin contact resistance, which manifests itself as the observed  peak at the onset of the superconducting transition.
Typically the spin accumulation at the SF interface is quantified by excess resistance, expressed as $\Delta R$ =
$\frac{P^{2}}{1-P^{2}} (\frac{\rho l_{sd}}{A})$, where P, $\rho$, and $l_{sd}$, are the spin polarization, resistivity,
and the spin diffusion length of the ferromagnet respectively, and A is the area of F/S junction \cite{jedema99}. This
expression cannot be used to quantify $\Delta R$ for a half metal as it diverges for $P=1$, but it is clear that for
half metals with $P$ close to 1 the spin
accumulation can be considerably larger than in other ferromagnets. \\
Spin accumulation leads to excess resistance, but that accumulation would occur is non-trivial. The zero-field state
can be supposed to generate triplets since the domain state of both ferromagnets can be considered as non-collinear.
Applying an in-plane field makes the F$_{1,2}$ magnetizations more collinear, but if the F' magnetization has a
component perpendicular to the interface the triplet magnetization axis would indeed be different from the bulk. In the
same vein, the effect would be less for the out-of-plane configuration, in which F' and F$_2$ are becoming more
collinear. Theoretical modeling will be needed to investigate this scenario. \\
To summarize, we demonstrated a triplet spin valve using a 100~$\%$ spin polarized ferromagnet and changing the field
from in-plane to out-of-plane, and found very large effects occurring up to quite high magnetic fields.  We also showed
that the interface transparency between the bulk magnet and the triplet-generating stack has a decided effect on the
efficiency of the TSV. Finally, a characteristic peak in the transition curve of the TSV with the field in plane was
explained in terms of spin accumulation caused by equal-spin Cooper pair breaking. We suggest that TSV's, in particular
those based on half metals, are good model systems for a systematical study of the parameters which are relevant for
triplet generation, although this still requires the development of a theoretical formalism based on half metals.

\begin{figure}[t]
\captionsetup{width=\columnwidth}
\begin{centering}
\includegraphics[width=6cm]{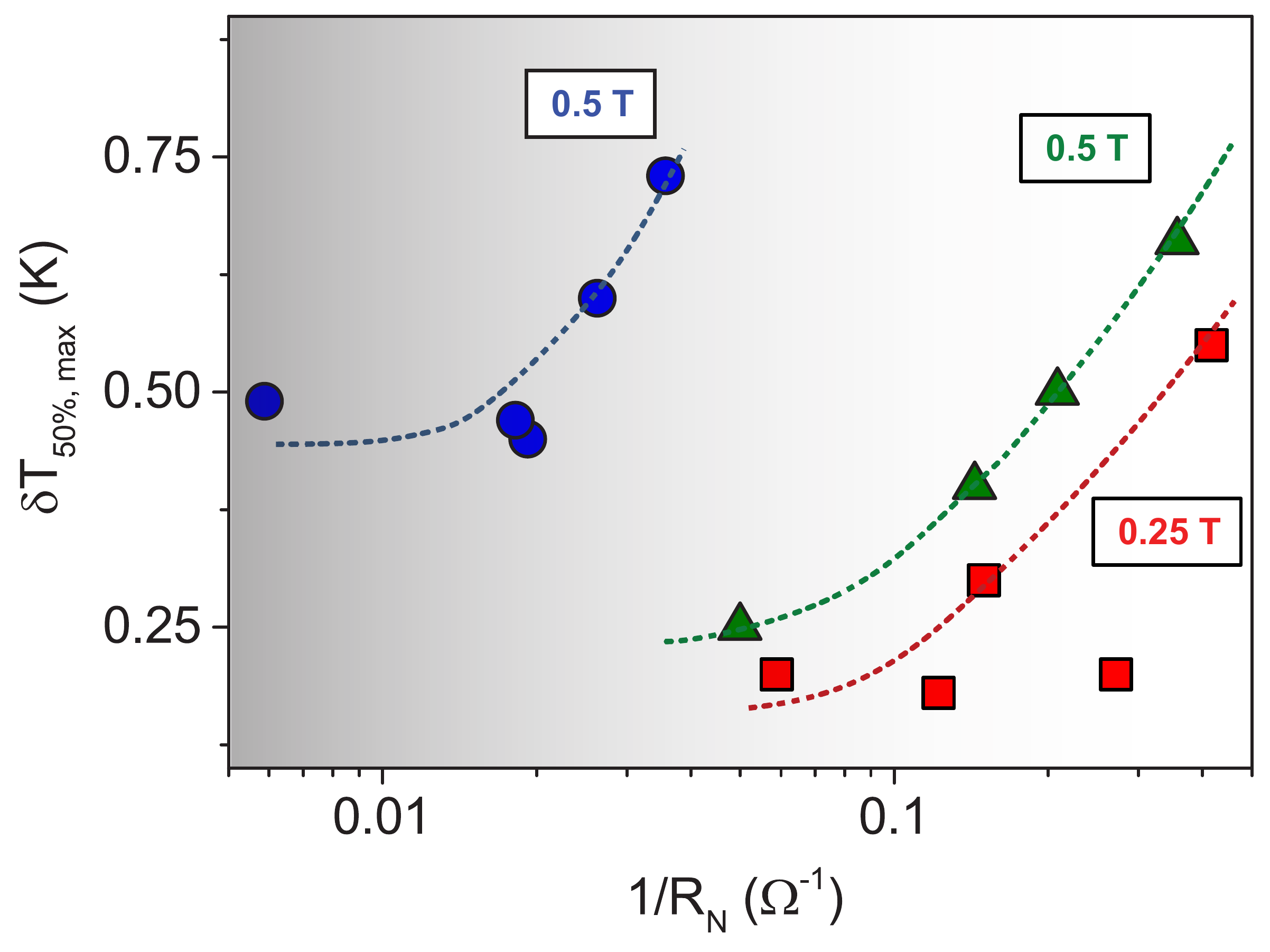}
\par\end{centering}
\caption{\textbf{Dependence of the triplet spin valve effect on the normal state resistance}. Variation of
T$_{50\%,max}$ plotted against the inverse of the normal state resistance $R_N$, for spin valve devices with
different interfaces between the CrO$_2$ layer and the Cu/Ni/MoGe stack. Blue circles: TSV25 (d$_S$ = 25~nm),
measured at 0.5~T; green triangles (red squares): TSV50 (d$_S$ = 50~nm), measured at 0.5~T (0.25~T).}
\end{figure}
\begin{figure}[tbh]
\captionsetup{width=\columnwidth}
\begin{centering}
\includegraphics[width=8cm]{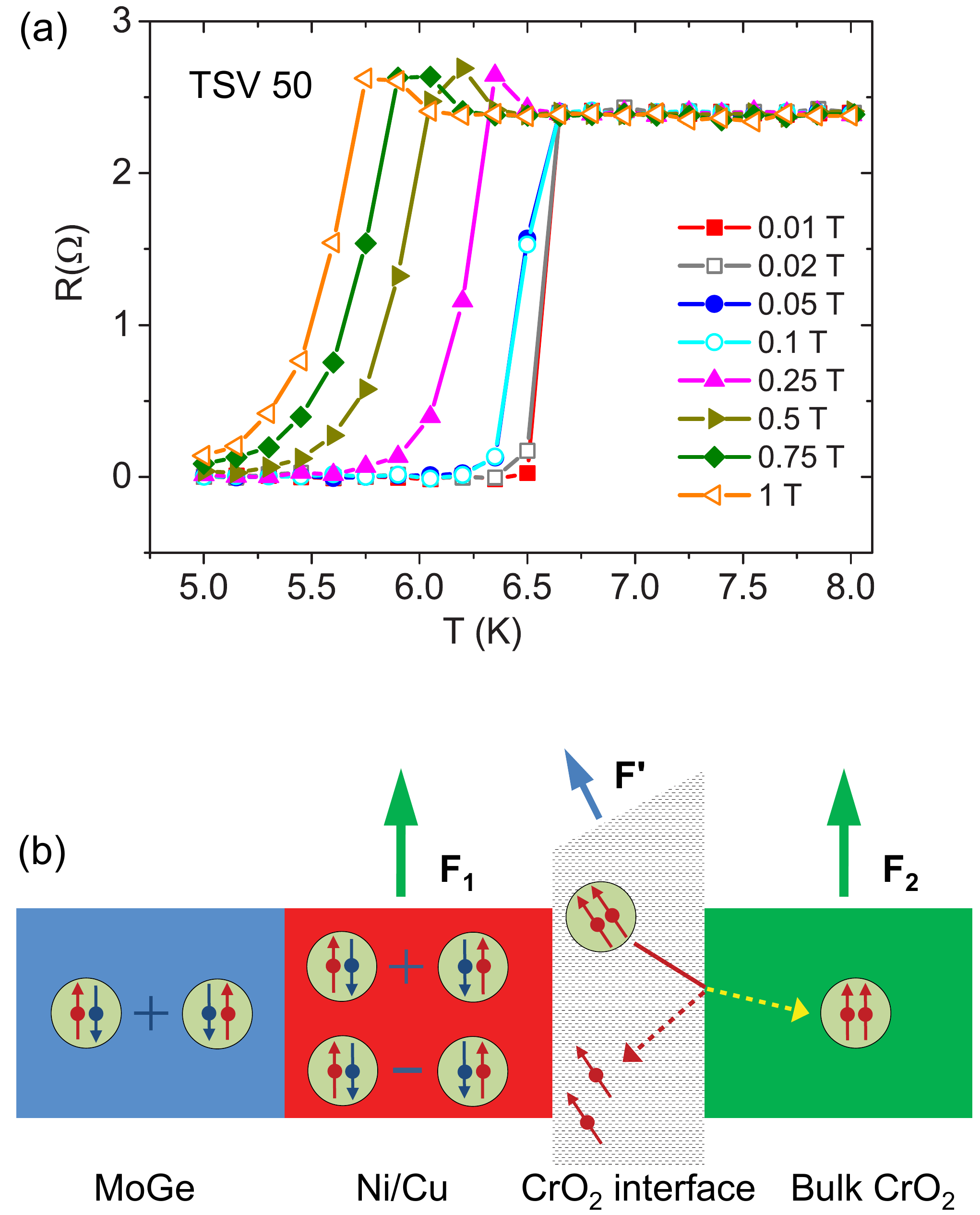}
\par\end{centering}
\caption{\textbf{Development of a resistance peak in a triplet spin valve with increasing in-plane field}
$\bf{a}$, Resistive transitions of a spin valve device TSV50, for different
values of the in-plane field between 0~T and 1~T. $\bf{b}$, pictorial representation of the effect of an extra
ferromagnetic layer F' between the mixer layer F$_1$ and the drainage layer F$_2$ on the generation of triplet pairs.
The green arrows represent magnetization directions of F$_{1,2}$) (in plane); the blue arrow indicates a interface layer F'
with magnetization direction out-of-plane.}
\end{figure}
\noindent \textbf{Methods}

Firstly, high quality epitaxial CrO$_{2}$ thin films were grown on TiO$_{2}$ substrates using chemical vapor deposition
in a two zone furnace. In one zone the substrate is kept at $390$$^{\circ}$C and the CrO$_{3}$ precursor is heated to
$260$$^{\circ}$C in the other zone. The precursor vapor is carried with O$_{2}$ gas (flow rate $100$ sccm) to the
substrate where it decomposes into CrO$_{2}$ over a very narrow temperature range ($390^{\circ}C-400^{\circ}C$) . In
order to fabricate triplet spin valves, a 10 $\mu$m wide bridge of trilayer (Cu/Ni/MoGe) was patterned on CrO$_{2}$
thin films via e-beam lithography followed by lift-off.

For electrical measurements, devices were connected in a 4-probe geometry to a PPMS chip holder, which was loaded on a
special rotator platform sample board for angle dependent magnetotransport measurements.

\noindent \textbf{Acknowledgements} Technical support from M.B.S. Hesselberth and D. Boltje, and discussions with S.
Bergeret are gratefully acknowledged. This work is part of the research programme of the Foundation for Fundamental
Research on Matter (FOM), which is part of the Netherlands Organisation for Scientific Research (NWO). It was supported
by a grant from the Leiden-Delft Consortium 'NanoFront' and by COST action MP1201). \\

\noindent \textbf{Author contributions} A.S. fabricated the samples, performed electrical measurements and analysed the
results. S.V. carried out part of the measurements and analysis. K. L. performed the magnetization measurements. J.A
supervised the work. All authors contributed to writing the manuscript and gave critical comments. \\

\noindent \textbf{Additional information} Supplementary information is available in the online version of the paper. \\

\noindent \textbf{Competing financial interests} The authors declare no competing financial interests. \\

\noindent \textbf{Corresponding Author} Correspondence to J. Aarts (aarts@physics.leidenuniv.nl)


\vspace{5mm}



\newpage

\begin{thebibliography}{10}
\expandafter\ifx\csname url\endcsname\relax
  \def\url#1{\texttt{#1}}\fi
\expandafter\ifx\csname urlprefix\endcsname\relax\def\urlprefix{URL }\fi
\providecommand{\bibinfo}[2]{#2}
\providecommand{\eprint}[2][]{\url{#2}}
%
\bibitem{fominov10} \bibinfo{author}{Fominov, Y.~V.} \emph{et~al.}
\newblock \bibinfo{title}{{Superconducting Triplet Spin Valve}}.
\newblock \emph{\bibinfo{journal}{{Jetp Lett.}}}
  \textbf{\bibinfo{volume}{{91}}}, \bibinfo{pages}{{308--313}}
  (\bibinfo{year}{{2010}}).
%
\bibitem{feofanov10} \bibinfo{author}{Feofanov, A.~K.} \emph{et~al.}
\newblock \bibinfo{title}{{Implementation of
  superconductor/ferromagnet/superconductor pi-shifters in superconducting
  digital and quantum circuits}}.
\newblock \emph{\bibinfo{journal}{{Nat. Phys.}}}
  \textbf{\bibinfo{volume}{{6}}}, \bibinfo{pages}{{593--597}}
  (\bibinfo{year}{{2010}}).
%
\bibitem{linder13} \bibinfo{author}{Enoksen, H.}, \bibinfo{author}{Linder, J.} \&
  \bibinfo{author}{Sudbo, A.}
\newblock \bibinfo{title}{{Pressure-induced 0-pi transitions and supercurrent
  crossover in antiferromagnetic weak links}}.
\newblock \emph{\bibinfo{journal}{{Phys. Rev. B}}}
  \textbf{\bibinfo{volume}{{88}}}, \bibinfo{pages}{214512-1--4} (\bibinfo{year}{{2013}}).
%
\bibitem{bergeret01} \bibinfo{author}{Bergeret, F.}, \bibinfo{author}{Volkov, A.} \&
  \bibinfo{author}{Efetov, K.}
\newblock \bibinfo{title}{{Long-range proximity effects in
  superconductor-ferromagnet structures}}.
\newblock \emph{\bibinfo{journal}{{Phys. Rev. Lett.}}}
  \textbf{\bibinfo{volume}{{86}}}, \bibinfo{pages}{{4096--4099}}
  (\bibinfo{year}{{2001}}).
%
\bibitem{kadigr01} \bibinfo{author}{Kadigrobov, A.}, \bibinfo{author}{Shekhter, R. I.} \&
  \bibinfo{author}{Jonson, M.}
\newblock \bibinfo{title}{Quantum spin fluctuations as a source of long-range
proximity effects in diffusive ferromagnet-superconductor structures}.
\newblock \emph{\bibinfo{journal}{{Europhys. Lett.}}}
  \textbf{\bibinfo{volume}{{54}}}, \bibinfo{pages}{{394--400}}
  (\bibinfo{year}{{2001}}).
%
\bibitem{bergeret03} \bibinfo{author}{Bergeret, F.}, \bibinfo{author}{Volkov, A.} \&
  \bibinfo{author}{Efetov, K.}
\newblock \bibinfo{title}{{Manifestation of triplet superconductivity in
  superconductor-ferromagnet structures}}.
\newblock \emph{\bibinfo{journal}{{Phys. Rev. B}}}
  \textbf{\bibinfo{volume}{{68}}}, \bibinfo{pages}{064513-1--11} (\bibinfo{year}{{2003}}).
%
\bibitem{houzet07} \bibinfo{author}{Houzet, M.} \& \bibinfo{author}{Buzdin, A.~I.}
\newblock \bibinfo{title}{{Long range triplet Josephson effect through a
  ferromagnetic trilayer}}.
\newblock \emph{\bibinfo{journal}{{Phys. Rev. B}}}
  \textbf{\bibinfo{volume}{{76}}}, \bibinfo{pages}{060504(R)-1--4} (\bibinfo{year}{{2007}}).
%
\bibitem{eschrig08} \bibinfo{author}{Eschrig, M.} \& \bibinfo{author}{Lofw\"{a}nder, T.}
\newblock \bibinfo{title}{{Triplet supercurrents in clean and disordered halfmetallic ferromagnets}}.
\newblock \emph{\bibinfo{journal}{{Nature Physics}}}
  \textbf{\bibinfo{volume}{{4}}} \bibinfo{pages}{{138--143}} (\bibinfo{year}{{2008}}).
%
\bibitem{khaire10} \bibinfo{author}{Khaire, T.~S.}, \bibinfo{author}{Khasawneh, M.~A.},
  \bibinfo{author}{Pratt, W.~P., Jr.} \& \bibinfo{author}{Birge, N.~O.}
\newblock \bibinfo{title}{{Observation of Spin-Triplet Superconductivity in
  Co-Based Josephson Junctions}}.
\newblock \emph{\bibinfo{journal}{{Phys. Rev. Lett.}}}
  \textbf{\bibinfo{volume}{{104}}}, \bibinfo{pages}{137002-1--4} (\bibinfo{year}{{2010}}).
%
\bibitem{robinson10} \bibinfo{author}{Robinson, J. W.~A.}, \bibinfo{author}{Witt, J. D.~S.} \&
  \bibinfo{author}{Blamire, M.~G.}
\newblock \bibinfo{title}{{Controlled Injection of Spin-Triplet Supercurrents
  into a Strong Ferromagnet}}.
\newblock \emph{\bibinfo{journal}{{Science}}} \textbf{\bibinfo{volume}{{329}}},
  \bibinfo{pages}{{59--61}} (\bibinfo{year}{{2010}}).
%
\bibitem{khasawneh12}  \bibinfo{author}{Khasawneh, M.~A.},
  \bibinfo{author}{Pratt, W.~P., Jr.} \& \bibinfo{author}{Birge, N.~O.}
\newblock \bibinfo{title}{{xx}}.
\newblock \emph{\bibinfo{journal}{{Sup. Sci. Techn.}}}
  \textbf{\bibinfo{volume}{{104}}}, \bibinfo{pages}{024005-1--7} (\bibinfo{year}{{2012}}).
%
\bibitem{anwar12} \bibinfo{author}{Anwar, M.~S.}, \bibinfo{author}{Veldhorst, M.},
  \bibinfo{author}{Brinkman, A.} \& \bibinfo{author}{Aarts, J.}
\newblock \bibinfo{title}{{Long range supercurrents in ferromagnetic CrO$_2$ using
  a multilayer contact structure}}.
\newblock \emph{\bibinfo{journal}{{Appl. Phys. Lett.}}}
  \textbf{\bibinfo{volume}{{100}}}, \bibinfo{pages}{052602-1--3} (\bibinfo{year}{{2012}}).
%
\bibitem{klose12} \bibinfo{author}{Klose, C.}, \bibinfo{author}{Khaire, T.~S.},
  \bibinfo{author}{Wang, Y.}, \bibinfo{author}{Pratt Jr., W.~P.} \& \bibinfo{author}{Birge, N.~O.},
  \bibinfo{author}{McMorran, B.~J.}, \bibinfo{author}{Ginley, T.~P.}, \bibinfo{author}{Borchers, J.~A.},
  \bibinfo{author}{Kirby, B.~J.}, \bibinfo{author}{Maranville, B.~B.} \& \bibinfo{author}{Unguris, J.}
\newblock \bibinfo{title}{{Optimization of Spin-Triplet Supercurrent in Ferromagnetic Josephson Junctions}}.
\newblock \emph{\bibinfo{journal}{{Phys. Rev. Lett.}}}
  \textbf{\bibinfo{volume}{{108}}}, \bibinfo{pages}{127002-1--5} (\bibinfo{year}{{2012}}).
%
\bibitem{leksin12} \bibinfo{author}{Leksin, P.~V.} \emph{et~al.}
\newblock \bibinfo{title}{{Evidence for Triplet Superconductivity in a
  Superconductor-Ferromagnet Spin Valve}}.
\newblock \emph{\bibinfo{journal}{{Phys. Rev. Lett.}}}
  \textbf{\bibinfo{volume}{{109}}}, \bibinfo{pages}{057005-1--4} (\bibinfo{year}{{2012}}).
%
\bibitem{wang14} \bibinfo{author}{Wang, X.~L.} \emph{et~al.}
\newblock \bibinfo{title}{{Giant triplet proximity effect in superconducting
  pseudo spin valves with engineered anisotropy}}.
\newblock \emph{\bibinfo{journal}{{Phys. Rev. B}}}
  \textbf{\bibinfo{volume}{{89}}}, \bibinfo{pages}{140508(R)-1--4} (\bibinfo{year}{{2014}}).
%
\bibitem{banerjee14} \bibinfo{author}{Banerjee, N.} \emph{et~al.}
\newblock \bibinfo{title}{{Evidence for spin selectivity of triplet pairs in
  superconducting spin valves}}.
\newblock \emph{\bibinfo{journal}{{Nat. Commun.}}}
  \textbf{\bibinfo{volume}{{5}}}, \bibinfo{pages}{048-1--6} (\bibinfo{year}{{2014}}).
%
\bibitem{jara14} \bibinfo{author}{Jara, A.~J.} \emph{et~al.}
\newblock \bibinfo{title}{{Angular dependence of superconductivity in superconductor/spin-valve heterostructures}}.
\newblock \emph{\bibinfo{journal}{{Phys. Rev. B}}}
  \textbf{\bibinfo{volume}{{5}}}, \bibinfo{pages}{184502-1--9} (\bibinfo{year}{{2014}}).
%
\bibitem{flokstra14} \bibinfo{author}{{Flokstra}, M.~G.} \emph{et~al.}
\newblock \bibinfo{title}{{Controlled suppression of superconductivity by the
  generation of polarized Cooper pairs in spin valve structures}}.
\newblock \emph{\bibinfo{journal}{{ArXiv e-prints}}}  (\bibinfo{year}{{2014}}).
\newblock \eprint{{1404.2950}}.
%
\bibitem{keizer06} \bibinfo{author}{Keizer, R.}, \bibinfo{author}{G\"{o}nnenwein, S. T. B.}, \bibinfo{author}{Klapwijk,
    T. M.}, \bibinfo{author}{Miao, G.}, \bibinfo{author}{Xiao, G.} \& \bibinfo{author}{Gupta, A.}
\newblock \bibinfo{title}{{A spin triplet supercurrent through the
  half-metallic ferromagnet CrO2}}.
\newblock \emph{\bibinfo{journal}{{Nature}}} \textbf{\bibinfo{volume}{{439}}},
  \bibinfo{pages}{{825--827}} (\bibinfo{year}{{2006}}).
%
\bibitem{anwar10} \bibinfo{author}{Anwar, M.~S.}, \bibinfo{author}{Czeschka, F.},
  \bibinfo{author}{Hesselberth, M.}, \bibinfo{author}{Porcu, M.} \&
  \bibinfo{author}{Aarts, J.}
\newblock \bibinfo{title}{{Long-range supercurrents through half-metallic
  ferromagnetic CrO2}}.
\newblock \emph{\bibinfo{journal}{{Phys. Rev. B}}}
  \textbf{\bibinfo{volume}{{82}}}, \bibinfo{pages}{100501(R)-1--4} (\bibinfo{year}{{2010}}).
%
\bibitem{fominov02} \bibinfo{author}{Fominov, Y.~V.}, \bibinfo{author}{Chtchelkatchev, N.~M.} \&
  \bibinfo{author}{Golubov, A.~A.}
\newblock \bibinfo{title}{Nonmonotonic critical temperature in
  superconductor/ferromagnet bilayers}.
\newblock \emph{\bibinfo{journal}{Phys. Rev. B}} \textbf{\bibinfo{volume}{66}},
  \bibinfo{pages}{014507-1--13} (\bibinfo{year}{2002}).
%
\bibitem{cirillo05} \bibinfo{author}{Cirillo, C.}, \bibinfo{author}{Prischepa, S.L.},
  \bibinfo{author}{Salvato, M. }, \bibinfo{author}{Attanasio, C.A.},
  \bibinfo{author}{Hesselberth, M.}, \& \bibinfo{author}{Aarts, J.}
\newblock \bibinfo{title}{Superconducting proximity effect and interface transparency in Nb/PdNi bilayers}.
\newblock \emph{\bibinfo{journal}{Phys. Rev. B}} \textbf{\bibinfo{volume}{72}},
  \bibinfo{pages}{144511-1--7} (\bibinfo{year}{2005}).
%
\bibitem{baarle03} \bibinfo{author}{Baarle, G. J. van}, \bibinfo{author}{Troianovski, A. M.},
  \bibinfo{author}{Nishizaki, T.}, \bibinfo{author}{Kes, P. H.} \& \bibinfo{author}{Aarts, J.}
\newblock \bibinfo{title}{{Imaging of vortex configurations in thin films by scanning-tunneling
microscopy}}.
\newblock \emph{\bibinfo{journal}{{Appl. Phys. Lett.}}}
  \textbf{\bibinfo{volume}{{82}}}, \bibinfo{pages}{1081-1084} (\bibinfo{year}{{2012}}).
%
\bibitem{jedema99} \bibinfo{author}{Jedema, F. J. }, \bibinfo{author}{van Wees, B. J.},
  \bibinfo{Hoving, B. H.}, \bibinfo{author}{Filip, A. T.} \& \bibinfo{author}{Klapwijk, T. M.}
\newblock \bibinfo{title}{Spin-accumulation-induced resistance in mesoscopic ferromagnet-superconductor junctions}.
\newblock \emph{\bibinfo{journal}{Phys. Rev. B}} \textbf{\bibinfo{volume}{60}},
  \bibinfo{pages}{16549-16552} (\bibinfo{year}{1999}).

\end{thebibliography}
\end{document}